\documentclass[aps,titlepage,12pt,superscriptaddress]{revtex4}
\usepackage{epsfig}
\usepackage{times}
\begin{document}

\title{Cooperation and competition between pair and multi-player social games in spatial populations}

\author{Attila Szolnoki}
\email{szolnoki.attila@energia.mta.hu}
\affiliation{Institute of Technical Physics and Materials Science, Centre for Energy Research, P.O. Box 49, H-1525 Budapest, Hungary}

\author{Xiaojie Chen}
\affiliation{School of Mathematical Sciences,
University of Electronic Science and Technology of China, Chengdu 611731, China}

\begin{abstract}
The conflict between individual and collective interests is in the heart of every social dilemmas established by evolutionary game theory. We cannot avoid these conflicts but sometimes we may choose which interaction framework to use as a battlefield. For instance some people like to be part of a larger group while other persons prefer to interact in a more personalized, individual way. Both attitudes can be formulated via appropriately chosen traditional games. In particular, the prisoner's dilemma game is based on pair interaction while the public goods game represents multi-point interactions of group members. To reveal the possible advantage of a certain attitude we extend these models by allowing players not simply to change their strategies but also let them to vary their attitudes for a higher individual income. We show that both attitudes could be the winner at a specific parameter value. Interestingly, however, the subtle interplay between different states may result in a counterintuitive evolutionary outcome where the increase of the multiplication factor of public goods game drives the population to a fully defector state. We point out that the accompanying pattern formation can only be understood via the multipoint or multi-player interactions of different microscopic states where the vicinity of a particular state may influence the relation of two other competitors.
\end{abstract}

\maketitle

\section*{Introduction}

Evolutionary game theory provides an efficient mathematical tool to describe the conflicts of interest between individuals or competitors \cite{maynard_82,sigmund_10,nowak_06}. This framework is proved to be extremely useful in broad variety of scientific disciplines. Examples can be given starting from the tiny scale of viruses or bacterias, via plants, animals till human societies \cite{elena_cov14,frey_11,schuster_bs11,camerer_s03,ostrom_90}. But complex and spatially structured interactions of companies or countries also raise critical problems, like climate change or depletion of natural resources, where the importance of this approach is indisputable \cite{milinski_pnas08,pacheco_plrev14,wang_x_rspa20,szolnoki_epl17,yang_lh_csf21,li_k_srep16,shao_yx_epl19,he_n_amc19,floria_pre09}.  
The mentioned conflicts can be described by appropriately chosen elements of a payoff matrix where the utility of a certain strategy depends on the strategy choice of the partner. In the so-called social dilemma games the actors may choose between two conceptually different strategies, called as cooperator and defector states. While the former invests some effort into the common interest the latter does not, but only enjoys the benefit of the other's contribution. Hence it pays individually not to cooperate, but if all actors behave rationally then an undesired end, frequently called as the tragedy of the commons, is inevitable \cite{hardin_g_s68}. Not surprisingly, due to the importance of the problems, we can witness a very intensive research activity in the last decades where scientists try to identify possible mechanisms, escape routes and conditions which promote cooperation and maintain a sustainable state for systems having limited resources \cite{richter_bs19,liu_rr_amc19,han_jrsif15,zhang_lm_epl19,fu_y_c21,pinheiro_rsos21,wang_sx_pla21,amaral_pre20,he_qp_epl20,xu_zj_c19,gracia-lazaro_csf13,blackmail1,blackmail2}.

According to the traditional approach the conflict can be handled in the form of pair interactions between competitors, or alternatively we assume multi-point interactions where actors benefit can only be calculated on the basis of simultaneous decisions of a smaller group. In social dilemmas the first description is used in prisoner's dilemma game or snow-drift games, while public goods game represents the latter approach \cite{szabo_pr07,perc_jrsi13}.

It is our everyday life experience that people are different and some of them like to participate in large scale joint ventures while others prefer to focus on face to face, hence limited or personalized interactions. To make a distinction between these postures we will call these preferences as ``alone" and ``together" attitudes respectively. Let us emphasize that in our belief the referred ``alone" direction is not equivalent to the so-called ``loner" strategy frequently used in evolutionary game theory \cite{hauert_s02,michor_n02,deng_zh_pa18,hu_kp_epl20}. While the latter expresses a situation when a player does not properly interact with anybody else, hence its payoff is independent of the strategy choice of other players. In our present case, however, what we call ``alone" attitude covers exactly the way how conflicts are treated for instance in the prisoner's dilemma game. Here an actor interacts with others individually, behaves as a defector or cooperator, and its payoff does depend on the strategy choice of the other. When a player represents ``together" attitude then multi-point or group-like interactions are preferred. This situation is grabbed by a public goods game where the simultaneous strategy decisions of group members determine the individual payoff of every related actors. It is also a reasonable assumption that the above described preferences are not necessarily fixed, but they may also change similarly to the applied strategies. This presumption establishes a simple four-state model where
both individual strategy and attitude co-evolve during a microscopic adoption. In this way we can study the vitality of different combination of strategies and attitudes in the framework of an evolutionary game model. As we will show, ``together" attitude can dominate the population if the external circumstances provide safety condition for the cooperator strategy. In other words, when the multiplication factor, which is the key parameter of public goods game, is large enough then only ``together" attitude survives. However, the general role of the mentioned parameter is less straightforward, because in some cases the system evolves into a full defection state when we increase the value of the multiplication factor. The explanation of this counterintuitive behavior is based on a delicate interaction of competing states where the vicinity of a third party may change the relation of competing states.

\section*{Methods}

For simplicity and for the shake of proper comparison with previous studies we assume that players are distributed on an $L \times L$ square grid where periodic boundary conditions are applied. The actual status of $i$ player is described by two labels. The first one determines whether player $i$ is cooperator ($C$) or defector ($D$). The second label describes whether player $i$ prefers pair or multipoint (multi-player) interactions. These preferences are denoted by $``A"$ (alone) and $``O"$ (together), respectively. Therefore the population can be described by a four-state model where every player is in $A_D$, $A_C$, $O_D$, or $O_C$ state. For example, an $A_D$ player plays prisoner's dilemma game with its neighbors exclusively and collects a $T$ temptation value from a cooperator neighbor, or gets nothing from a defector player. Similarly, an $A_C$ player also plays prisoner's dilemma game with its neighbors and collects $R=1$ payoff from a cooperator neighbor while gets nothing from a defector partner. Shortly, we apply the weak prisoner's dilemma game parametrization where $R=1$, $S=0$, and $P=0$ are fixed, while $T>1$ is the control parameter characterizing the strength of the dilemma.

\begin{figure}[ht]
\centering
\includegraphics[width=10cm]{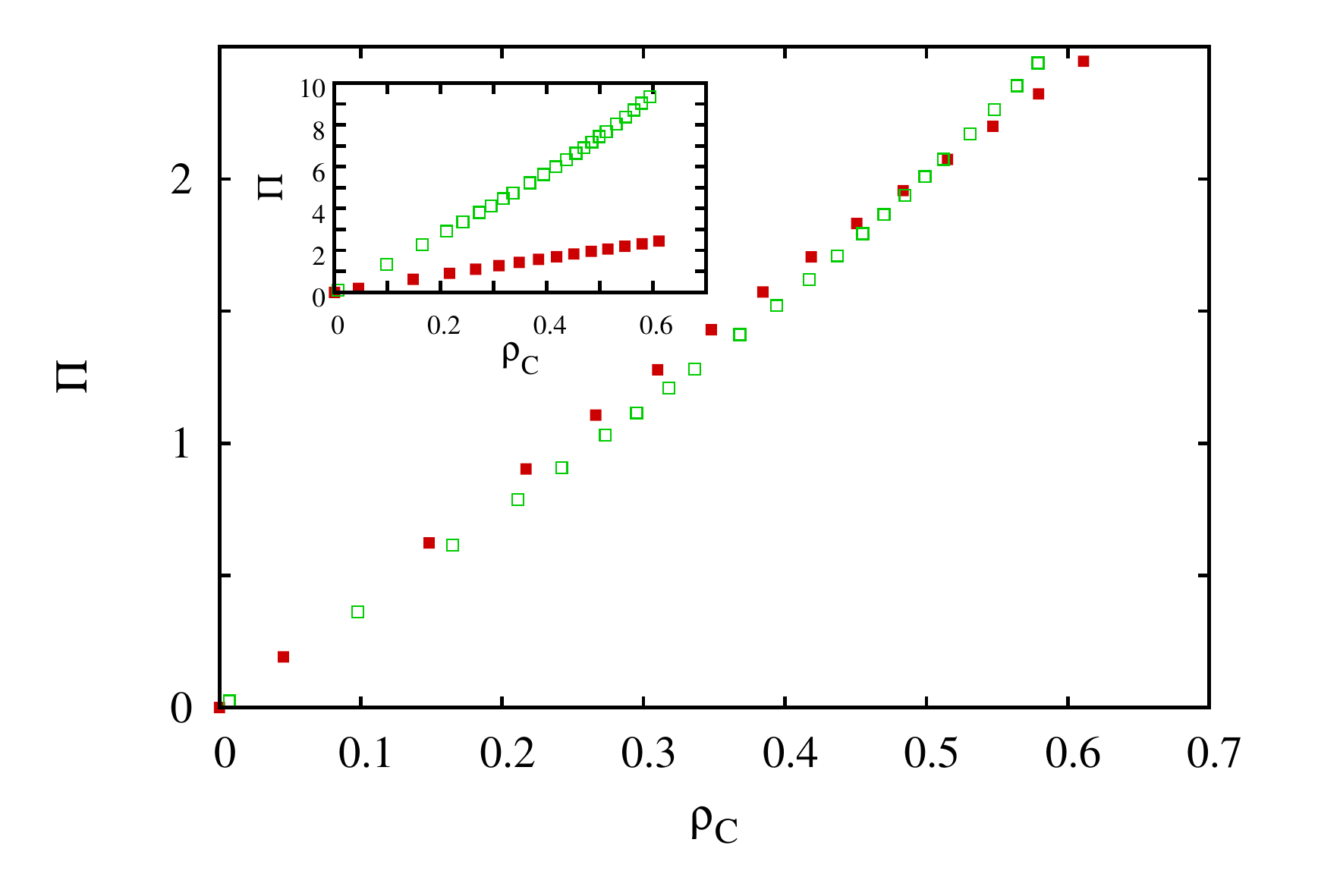}
\caption{Average $\Pi$ payoff values in dependence of cooperation level $\rho_C$ for prisoner's dilemma game (filled red box) and public goods game (open green box). Inset shows the proper values of the original games, while the main plot shows the case when the payoff values of public goods game are rescaled by an $\alpha=0.266$ factor. In this case the accumulated difference between the payoff values of the games is minimal. The usage of scaled payoff vales makes possible the proper competition between the $``A"$ and $``O"$ attitudes allocated to the mentioned games.}
\label{scaling}
\end{figure}

When a player $i$ is in $O_C$ state then it would like to participate in every group ventures where it might be involved. Optimally, when $i$ is surrounded by $``O"$ players exclusively, player $i$ participates in five public goods games on a square grid. These games are organized by itself and by its neighbors who have also $``O"$ attitude. But of course, such a joint venture cannot be forced to a player with $``A"$ attitude, therefore the mentioned player $i$ must play a prisoner's dilemma game in the latter case. In other words, in an extreme case a player with $``O"$ attitude might only play prisoner's dilemma games with its neighbors if it is surrounded by $``A"$ attitude players exclusively. The definition of the public goods game agrees with the traditional definition \cite{perc_jrsi13}. Namely, the payoff collected by an $O_C$ player from a specific game is $\Pi = (r \cdot n_{O_C})/ n_{O} - 1$, where $n_{O_C}$ is the number of $O_C$ players in the group and $n_{O}$ denotes the number of players in the group who have $``O"$ attitude. Naturally, the minimal value of $n_{O}$ is 2 and it goes up to 5 in the optimal case we mentioned earlier. In case of an $O_D$ player the mentioned payoff is $\Pi = (r \cdot n_{O_C})/ n_{O}$, because this player does not contribute to the common pool.

\begin{figure}[ht]
\centering
\includegraphics[width=10cm]{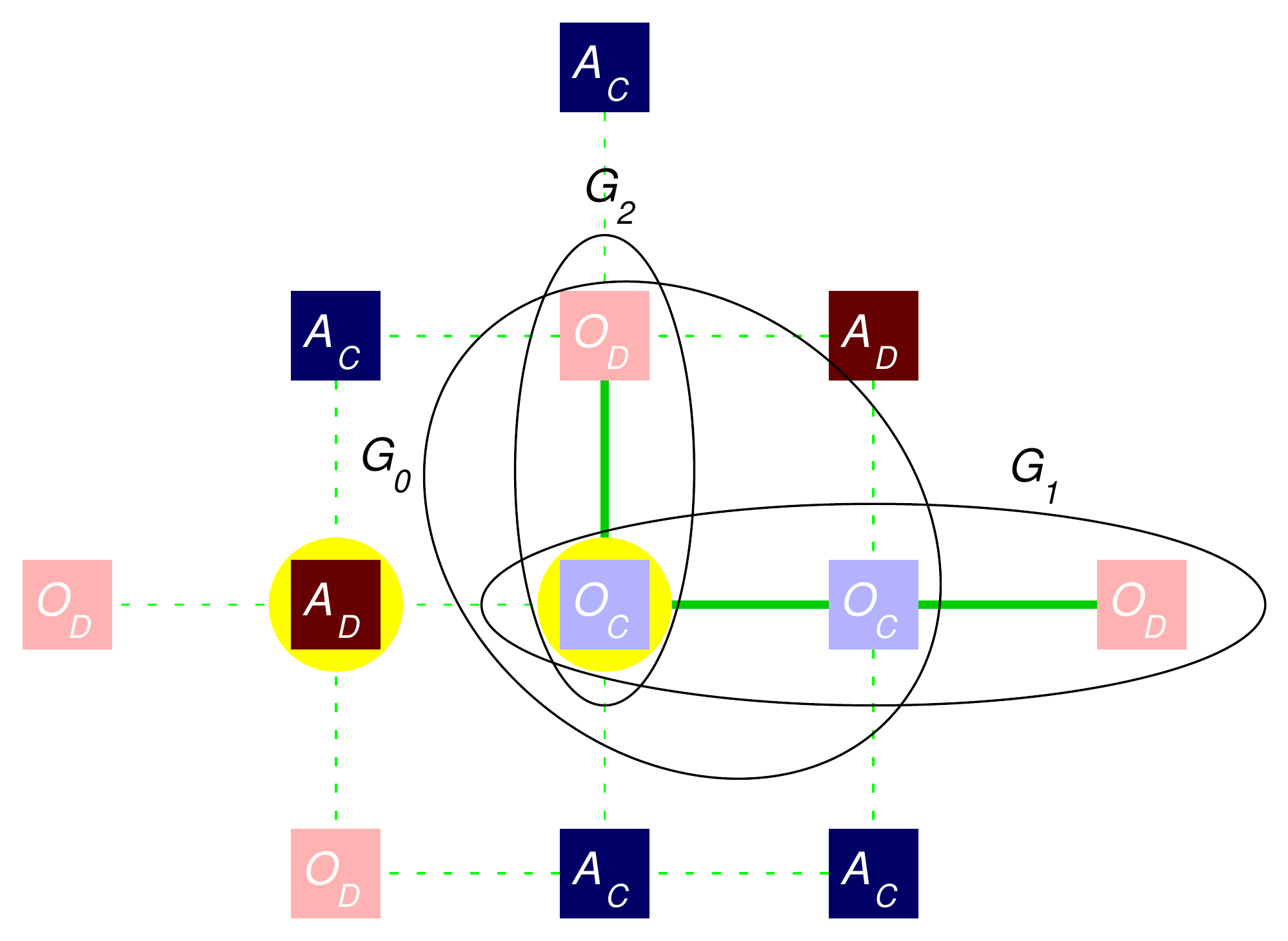}
\caption{Model setup to study the competition of ``alone" and ``together" attitudes. When calculating the payoff of $A_D$ marked by a yellow background then we sum the payoff elements of four prisoner's dilemma games played with the neighbors. Accordingly, $\Pi = 2 \cdot T + 2 \cdot P$ for the mentioned player. When we calculate the payoff of $O_C$ player also marked by yellow background then we should consider the fact that two of its neighbors have ``alone" attitude, hence $O_C$ should play prisoner's dilemma game with them. The remaining two neighbors have ``together" attitude hence the focal player can play a three-member public goods game with them in the marked $G_0$ group. The focal player also participates in a three-member public goods game of $G_1$ group and in a two-member game of $G_2$ group. For the shake of comparable payoff values we multiply the incomes of public goods games by an $\alpha$ factor. In sum, the total payoff of the mentioned $O_C$ player is $\Pi = S + R + \alpha \cdot (\pi_{G_0} + \pi_{G_1} +\pi_{G2})$ = $S+R+\alpha \cdot (r \cdot 2/3 - 1 + r \cdot 2/3 - 1 + r \cdot 1/2 - 1)$. The interactions where prisoner's dilemma game is used are marked by dashed while the interactions where public goods game is used are denoted by solid green links.}
\label{model}
\end{figure}

We must note, however, that the traditional prisoner's dilemma game and the public goods game were defined independently earlier therefore their average payoff values are not comparable. This is illustrated in the inset of Fig.~\ref{scaling} where we plotted the average payoff values in dependence of the average cooperation level for the classic games. Not really surprisingly, the average income in the public goods game is significantly higher because players can collect income from five games on a square grid. Therefore, to make the payoff values comparable and establish a proper competition between attitudes, we should scale the income of public goods game by an $\alpha$ factor. By using an appropriately chosen value the mentioned payoff values become comparable as the main plot of Fig.~\ref{scaling} demonstrates. In the rest of this work we will apply the proposed $\alpha=0.266$ weighting factor, but emphasize that our qualitative observations remain intact even if we use the original payoff values where $\alpha$ is formally equal to 1.

To summarize our model definition we plotted a specific configuration in Fig.~\ref{model} where we calculated the payoff values of the marked $A_D$ and $O_C$ players. As we noted, a player with an $``A"$ attitude always plays prisoner's dilemma games with its neighbors even if the latter players may represent alternative attitude. Accordingly, the total income of the mentioned $A_C$ players come from four values (where two of them are zero due to the standard $P=0$ parametrization of prisoner's dilemma game). Since $O_C$ player also has two neighbors who represent $``A"$ attitude therefore two parts of $O_C$'s total payoff come from prisoner's dilemma games, too. The remaining three payoff elements are the results of three traditional public goods games where the focal $G_0$ game contains 3 members and the focal player also participates in a three-member ($G_1$) and in a two-member ($G_2$) game organized by the neighbors. Importantly, the last three values are multiplied by the above mentioned $\alpha$ weighting factor.

The microscopic evolutionary dynamics is based on the payoff difference of the neighboring source and target players. More precisely, a player $i$ adopts the strategy $s_j$ from player $j$ with the probability
\begin{equation}
W = \frac{1}{1+\exp[(\Pi_i-\Pi_j)/K]} \, ,
\label{adopt}
\end{equation}
where $K$ quantifies the noise of imitation process \cite{szabo_pre98}. Importantly, players can also adopt an attitude from each other where they use the same imitation probability defined by Eq.~\ref{adopt}. These processes, however, are independent from each others, otherwise we may observe artificial consequence of coupled imitation processes. At the beginning of the Monte Carlo (MC) simulations we distribute both $C$, $D$ strategies and $``A"$, $``O"$ attitudes randomly among players who are on a square grid. During an elementary process we randomly select a target player $i$ and a neighboring player $j$. When the corresponding payoff values are calculated then we determine the imitation probability defined by Eq.~\ref{adopt}. After target player adopts the strategy and/or attitude of source player with the imitation probability. This step is repeated $L \times L$ times, which establishes a MC step. In our simulations we typically applied $L=600$ system size and 50000 MC steps to reach the stationary state, but in the vicinity of transition points we checked our data to avoid finite-size problems. To obtain comparable results with previous studies \cite{szabo_pre05,szolnoki_pre09c} we used $K=0.5$ noise level, but we have checked other values to confirm the robustness of our observations.

In the following we explore how different combination of strategies and attitudes compete and interact where the main parameters are the control parameters of the original games, that is the $T$ temptation to defect for the prisoner's dilemma and the $r$ multiplication factor for the public goods game.

\section*{Results}

Before presenting the evolutionary outcomes of spatially structured populations we note that in a well-mixed system cooperators cannot survive if $T>1$ in the prisoner's dilemma game or if $r<G$ in the public goods game where $G$ denotes the actual size of the group. On square lattice the critical temptation value is $T=1.0655$ where cooperators die out in the homogeneous prisoner's dilemma game and the minimal multiplication factor is $r=3.74$ which is necessary for cooperators to survive in the public goods game \cite{szabo_pre05,szolnoki_epl10}. Importantly, if we apply the earlier mentioned $\alpha=0.266$ weighting factor for the public goods payoff values then this critical $r$ value shifts to $r=4.33$. 

\begin{figure}[ht]
\centering
\includegraphics[width=12cm]{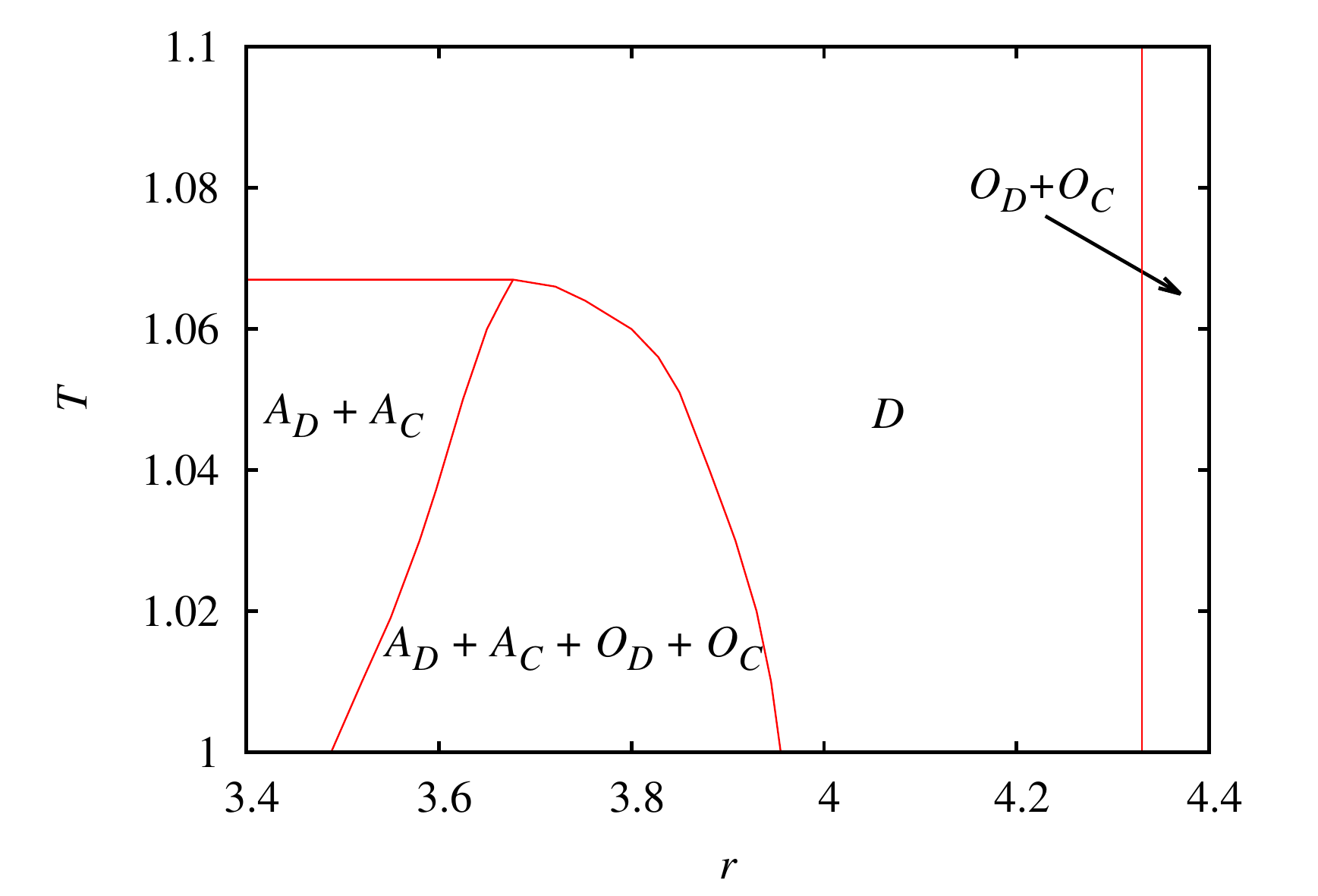}
\caption{Phase diagram of the four-state model in dependence of the key parameters of basic games. $r$ denotes the multiplication factor of the public goods game, while $T$ is the temptation to defect for the prisoner's dilemma game. At low $r$ values those states which represent pair interactions prevail. As we increase $r$, both attitudes, hence pair and multi-point interactions, coexist. Interestingly, as we support cooperation by increasing $r$ further the system evolves onto a full defection state. Cooperation can only recovered for very high $r$ values when the traditional public goods game with multi-point interactions prevail.}
\label{phase_diag}
\end{figure}

When we extend our traditional models and allow them to compete for space, the resulting behavior is more complicated, which is summarized in a phase diagram shown in Fig.~\ref{phase_diag}. Here we present the evolutionary outcomes on the plain of $T-r$ key parameters. When the multiplication factor $r$ is small then the system evolves into the traditional prisoner's dilemma game where only those players are present who prefer to play a game via pair interactions. In this case only the $T$ temptation value matters and if it is not too large then cooperators can coexist with defectors. If we increase the value of multiplication factor then a new kind of solution emerges: both types of attitudes survive in a complex pattern. Interestingly, as we will point out later, the larger multiplication factor does not involve increased cooperation level and higher well-being for players. Actually, the opposite is true. What is more, by increasing $r$ further the population evolves into a full defection state! And this sad destination can only be avoided if $r$ becomes significantly higher. In the latter case, however, our extended system evolves into the traditional model of public goods game where only players with $``O"$ attitude are present.

\begin{figure}[ht]
\centering
\includegraphics[width=12cm]{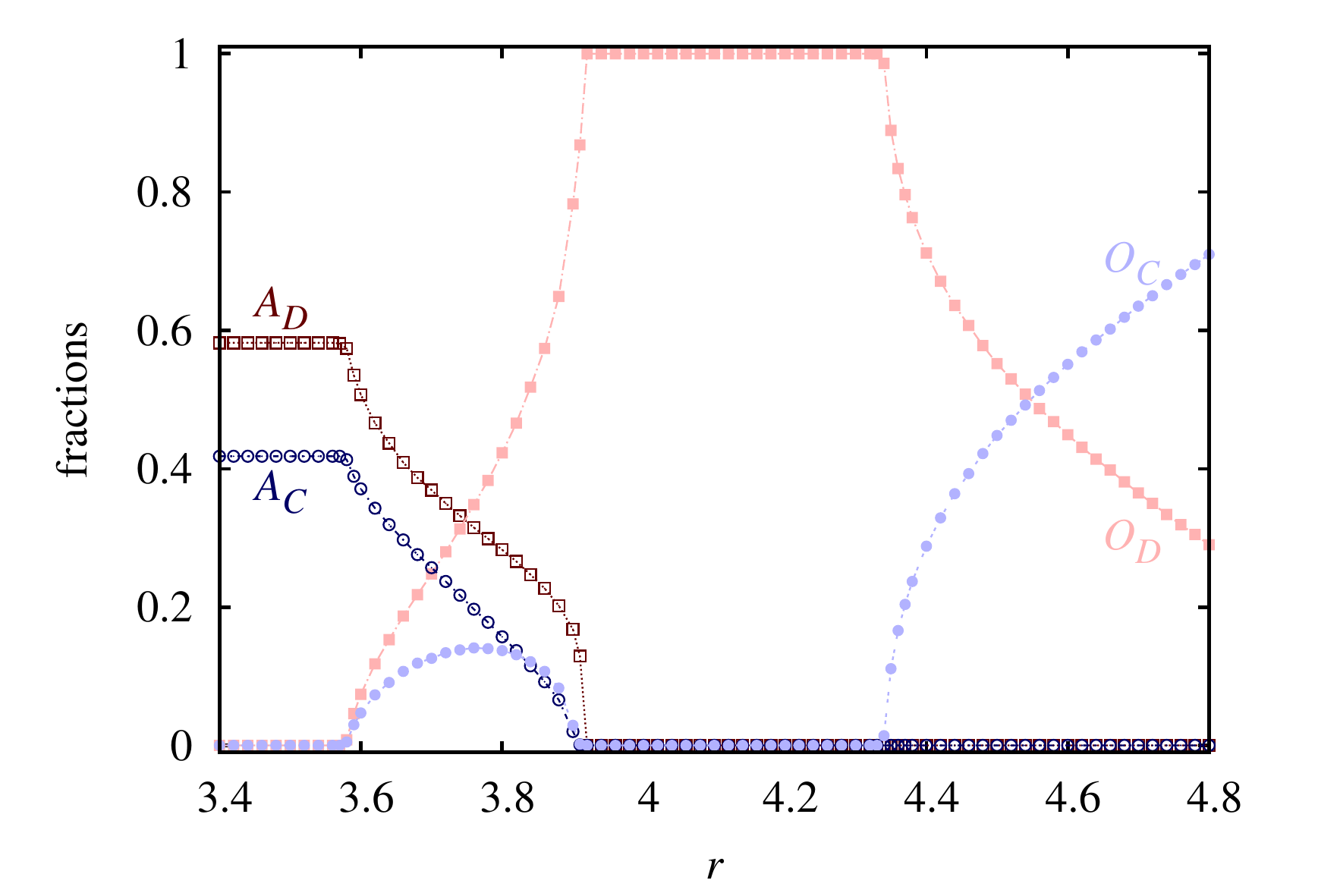}
\caption{Fractions of microscopic states in dependence of multiplication factor $r$ at temptation value $T=1.03$. The cross section of the phase diagram illustrates nicely that there are continuous phase transitions between different phases. In the coexistence phase between $3.57<r<3.92$ the fraction of $O_C$ state increases first while the portions of $A_D$ and $A_C$ decrease simultaneously. Above a critical $r$ value $O_C$ starts falling, which eventually leads the population to a full defector state.}
\label{cross}
\end{figure}

Figure~\ref{cross} helps to understand this highly counterintuitive system behavior where we present the stationary factions of all microscopic states in dependence of the multiplication factor. This plot suggests that the portion of $O_C$ state increases first when we enter the 4-state phase but above a critical $r$ value it starts falling again and diminishes when we reach the full defection state. This non-monotonous $r$-dependence of $O_C$ portion is the key feature which helps to understand the main mechanism that is responsible for the unexpected system reaction on enlarged $r$ parameter. It is important to note that $A_C$ and $O_C$ players can coexist with each other in the absence of the other two states when $r$ is small. It is because $O_C$ players cannot really enjoy the advantage of $``O"$ attitude for $r$ values. But this relation changes qualitatively when $r$ exceeds a certain value. In the latter case neighboring $O_C$ players can support each other so effectively that they can dominate $A_C$ players. As a result, $A_C$ and $O_C$ players cannot coexist anymore but the latter state prevails. The value of this threshold $r$ value agrees qualitatively with the position where the portion of $O_C$ starts decaying again. Perhaps this argument may sound unreasonable because we would expect better conditions for $O_C$ for higher $r$ values. But it is true that the success of $O_C$ over $A_C$ involves the former cooperators global decline. Namely, the value of $r$ is still not large enough for $O_C$ players to fight effectively against $O_D$ players. As we mentioned earlier, $r>4.33$ is necessary for stable coexistence in the pure public goods game. Hence $O_C$ players, who beat $A_C$ fellows, remain alone against $O_D$, and become vulnerable. In other words, the active presence of $A_C$ players is essential for $O_C$ players to survive at such a small $r$ value.

\begin{figure}[h!]
\centering
\includegraphics[width=12cm]{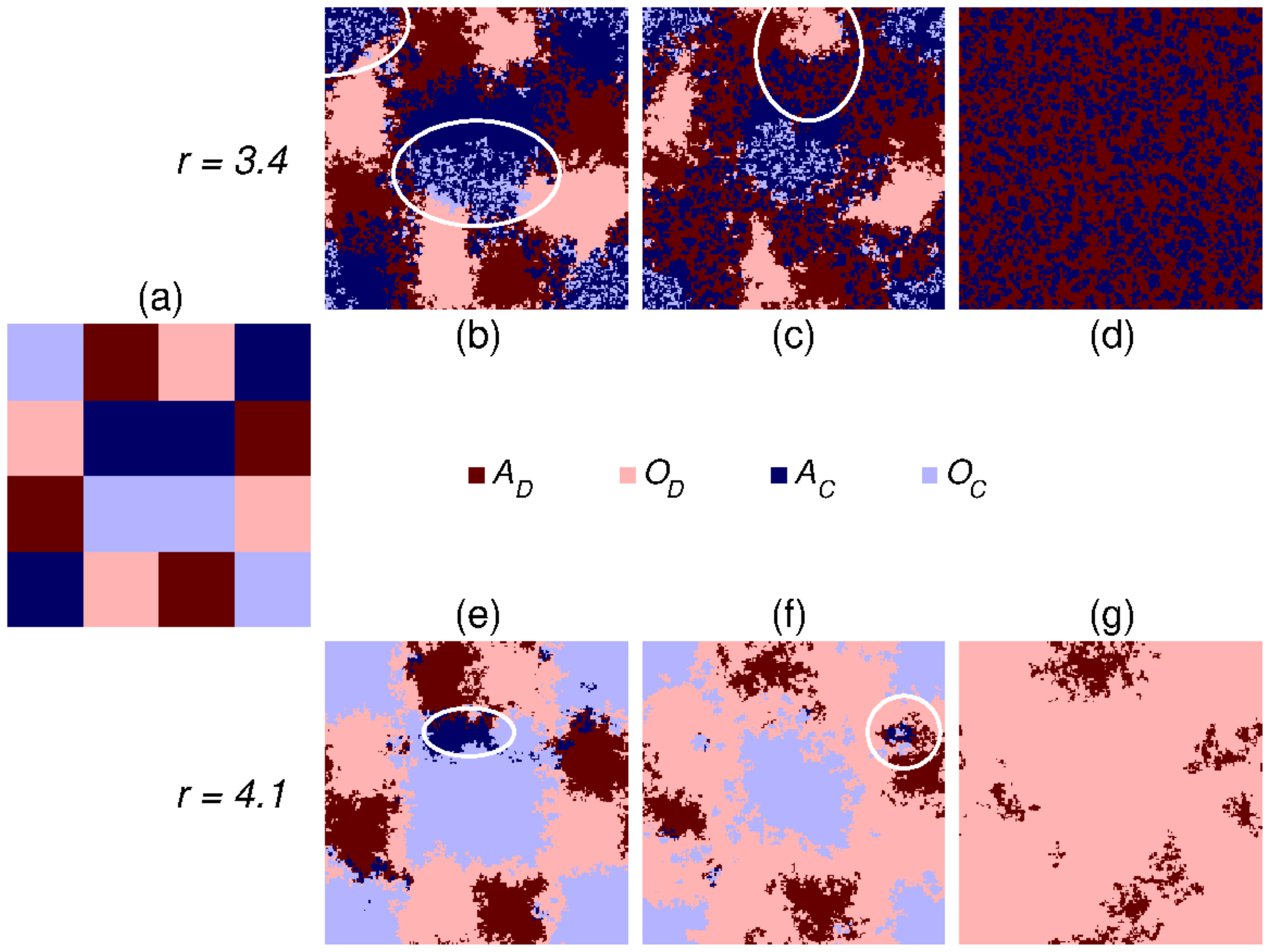}
\caption{Time evolution of the pattern formation at two representative values of multiplication factor $r$ when $T=1.03$. The common prepared initial state is shown in the left side where we arranged homogeneous patches of competing states denoted by different colors. As in previous plots, dark colors mark players having $``A"$ attitude, wile light colors denote players with $``O"$ attitude. Furthermore different shades of red color mark defector strategies, while various blue colors for cooperator strategies. The legend for the microscopic states are shown in the middle row of the plot. The top row from panels~(b) to (d) illustrates how the initial patchwork evolves at small ($r=3.4$) multiplication factor. Here, as white ellipses shown in panel~(b), $A_C$ and $O_C$ players can coexist due to the low value of $r$. Because of the modest temptation value $A_D$ and $A_C$ states can coexist and this mixed phase gradually beat $O_D$ players. This process is marked by a white ellipse in panel~(c). In the final state only players with $``A"$ attitude can survive. When $r$ is higher, as shown in the bottom row, $O_C$ beats $A_C$ who gradually diminishes. This is shown by a white ellipse in panel~(e). The lonely $O_C$, however, becomes vulnerable against $O_D$ and the previously victorious domains of $O_C$ players gradually shrink. In the absence of other states $A_D$ and $O_D$ would be neutral, but the vicinity of cooperators helps the latter to beat the former defector state. This is shown by a white ellipse in panel~(f). Finally, only defector state survive, but a slow voter-like stochastic dynamics will drive the system onto a homogeneous state.}
\label{snapshots}
\end{figure}

To confirm our argument in Fig.~\ref{snapshots} we present two representative pattern formation processes at two different values of $r$, which highlight the crucial role of $O_C$ players. For a clear demonstration of leading mechanisms we here used a prepared initial state where different microscopic states are arranged onto homogeneous patches \cite{szolnoki_pre11,szolnoki_pre11b}. This pattern is shown in the left column of Fig.~\ref{snapshots}. For a proper comparison we apply the same initial state for small and higher values of multiplication factor, too. When $r$ is small, shown in the top row, light blue $O_C$ players can coexist with dark blue $A_C$ players in the absence of the other two states. This solution is marked by white ellipses in panel~(b). We should not forget, however, that dark blue $A_C$ and dark red $A_D$ players can coexist because of the relatively low value of temptation $T$. This solution gradually displaces the light red $O_D$ domains as shown by an ellipse in panel~(c). Since $O_D$ cannot exploit $A_C$ players better than $A_D$ state therefore $O_D$ has no advantage over $A_D$, hence the random change will gradually destroy the homogeneous domain of $O_D$ state. Similarly, the mixture of $A_C$ and $O_C$ cannot be stable against the mentioned $A_C + A_D$ composition because a full cooperator state is unstable at these parameter values. As a result, the system eventually evolves into the traditional prisoner's dilemma game where only players with $``A"$ attitude are present. This final state is shown in panel~(d).

The relation of $O_C$ and $A_C$ players is qualitatively different at higher $r$ value, which has a significant consequence on the final outcome. This process is shown in the bottom row. Due to the higher $r$ value $O_C$ players can support each other more effectively and they gradually crowd out $A_C$ players. This process is illustrated by a declining island of dark blue $A_C$ state and marked by an ellipse in panel~(e). But light blue $O_C$ cannot be happy too long because they become vulnerable against light red $O_D$ players. We should note that the value of $r$ is still too small to ensure a stable coexistence between different players with $``O"$ attitude. Consequently, the fat light blue $O_C$ domain starts shrinking, as it is illustrated by panel~(f). Notably, defectors with different attitudes are neutral to each other: in the absence of cooperators there is no one to exploit, hence zero payoff difference in Eq.~\ref{adopt} dictates a random adoption of an attitude between defector players. There is however, a slight asymmetry which gives an advantage to $O_D$: when cooperators are still alive then their vicinity offers a higher propagation chance to $O_D$. Note that $A_C$ or $O_C$ would coexist with $A_D$ because they should play the traditional prisoner's dilemma game with the latter player. But $O_D$ propagates more efficiently in cooperator domains which results in a higher portion of $O_D$ when only defectors remain. This stage is shown in panel~(g).  We stress that, however, this is not the final state because the surface-tension free voter-model like coarsening will result in a homogeneous final destination \cite{cox_ap86}. Indeed, this is a logarithmically slow process where the attitude which has larger initial are has a higher chance to prevail \cite{dornic_prl01,helbing_pre10c}.

\begin{figure}[h!]
\centering
\includegraphics[width=12cm]{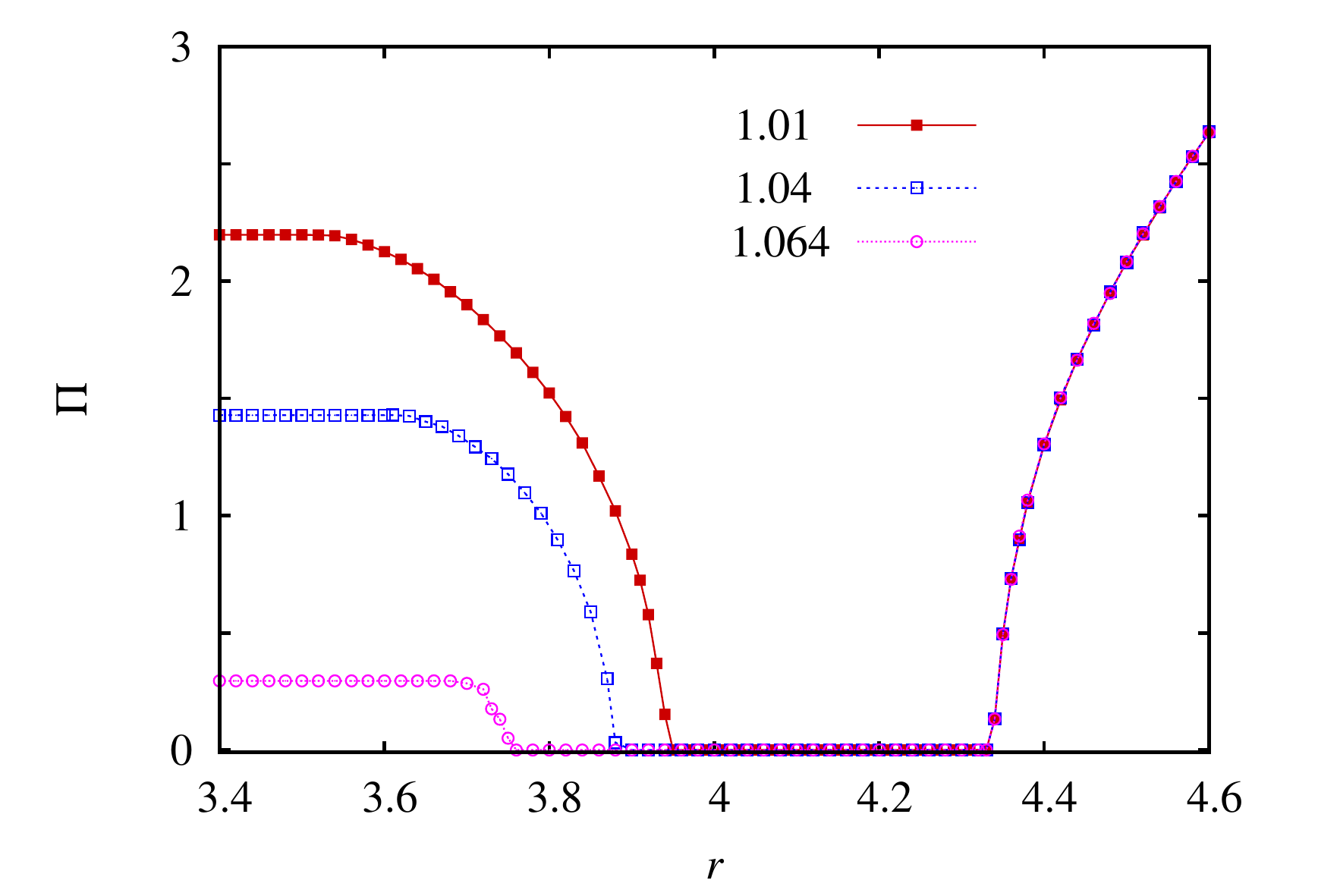}
\caption{The average income for the whole population in dependence of multiplication factor $r$ at different temptation values. The applied $T$ values are marked in the legend. $\Pi$ is constant for small $r$ values because only $``A"$ players, who play prisoner's dilemma game, are present. Interestingly, the emergence of ``together" attitude is detrimental for small $r$ values, because the 4-state solution provides a lower general income for players. This undesired consequence can only be fixed for high $r$ values when the traditional public goods game represented by $``O"$ players becomes dominant.}
\label{payoff}
\end{figure}

Summing up, the competition of the ``alone" and ``together" attitudes depends sensitively on the relationship of $O_C$ and $A_C$ players. When the multiplication factor of public goods game is low then these players can coexist, which prevents $O_D$ players to spread fast. As a result, the stability of players with ``alone" attitude can manifest and the final state is characterized by the coexistence of $A_C$ and $A_D$ players. But if we support $O_C$ players by increasing the multiplication factor then we may reach an undesired evolutionary outcome. Indeed, $O_C$ can dominate $A_C$ in the latter case, but it becomes vulnerable against $O_D$ players hence defection can prevail. In between the mentioned solutions there is a parameter region where all the assumed microscopic states coexist.

Interestingly, however, the presence of the ''together" attitude at low $r$ values does not really serve the population. This anomaly can already be seen in Fig.~\ref{cross} where the total fraction of cooperator states cannot be elevated by increasing $r$ in the 4-state phase. But this counter-intuitive behavior can be made more visible if we present the average general income of players in dependence of the multiplication factor. This is shown in Fig.~\ref{payoff} for different temptation values. These curves underline the unexpected consequence of the emerging $``O"$ attitude more dramatically: in sharp contrast to all previous experiences, the increase of $r$ results in a {\it lower} general income of players and drives the population toward an undesired evolutionary outcome. And this solution can only be avoided for even higher $r$ values when the system evolves into the two-state version of the traditional public goods game. 

\section*{Discussion}

Is it a generally valid recommendation to be part of a group or sometimes is it better to fight alone when we face a conflict that is represented by a social dilemma? Motivated by this question we here studied an extended model system where both attitudes are present beside the traditional defector and cooperator strategies. In particular, some players prefer to interact individually via pair interactions, hence they are tagged by an ``alone" ($``A"$) attitude, but others favor multi-point interactions by having a ``together" ($``O"$) character. While the original conflict between individual and collective interests is preserved, these preferences can be handled via a traditional prisoner's dilemma and a public goods game respectively. Our principal goal was to explore how these attitudes interact if we allow them to change via the widely applied imitation process driven by individual payoff difference of source and target players.

We note that the simultaneous usage of different games was already used by some previous works \cite{szolnoki_epl14,wang_z_pre14b}. But in these cases the application of diverse payoff elements was a predefined  microscopic rule which was assigned to specific players or the usage of alternative payoff elements were the subject of a time-course \cite{szolnoki_srep19,han_y_pa20,li_zb_amc19,hu_kp_pa19,wang_xj_pla20}. In our present case, however, the application of different payoff elements is the consequence of a natural evolutionary process which is driven by an individual imitation procedure.

In the extended system the two classical games emerge as extreme cases at different edges of the parameter range and their simultaneous presence are detected at an intermediate parameter interval. Interestingly, this mixed state, which evolved via individual success of strategies and attitudes, offers a lower general income for the whole population. This phenomenon and the resulting pattern formation can be explained as a subtle interplay between cooperator strategies of different individual attitudes. Importantly, the spatial character of population has a key function in the emergence of these solutions, because the vicinity of a third-party has a crucial role in how the relation of competing microscopic states may change and alter the evolutionary path \cite{szolnoki_pre14c,kelsic_n15,szolnoki_njp15}. This feature can be detected in several alternative systems and will always inspire future studies of networked or spatially organized populations.

Interestingly, our extended model offers cooperator strategy to survive for smaller synergy factor values which cannot be observed in the traditional model where only public goods game is considered. For a fair competition of attitudes represented by different games, we applied scaled, or weighted values for public goods game to ensure comparable incomes for both games. Interestingly, our qualitative observations remain intact when we apply the original unweighted payoff values of the public goods game. In the latter case the only difference is the parameter range where the 4-state solution emerges moves toward smaller $r$ values and evidently the size of the coexistence phase shrinks significantly. For example for $T=1.02$ this interval is between $2.16<r<2.28$. In sum, our observations fit nicely into previous research experiences which highlight that multi-state spatial population offer novel system behaviors that cannot be detected in well-mixed or unstructured systems and new kind of solutions emerge which would not exist otherwise \cite{chen_xj_epl10,dobramysl_jpa18,chen_xj_srep16,amaral_rspa20,brown_pre19,szolnoki_epl15}.

\section*{Acknowledgements}

This research was supported by the National Natural Science Foundation of China (Grants Nos. 61976048 and 61503062) and the Fundamental Research Funds of the Central Universities of China.

\section*{Author contributions statement}

A. S. and X. C. designed and performed the research as well as wrote the paper.

\section*{Additional information}

\textbf{Competing interests} The authors declare no competing interests.. 


\end{document}